\RequirePackage{etoolbox}
\csdef{input@path}{{style/}{graphics/}}
\documentclass[aoas,MSNbibl,nameyear,dvips]{arximspdf}
\makeatletter
   \@ifpackageloaded{graphicx}{}{\usepackage{graphicx}}
\makeatother
\usepackage{mathbh}
\usepackage{graphicx}

\doi{10.1214/14-AOAS789} 
\volume{9}
\issue{1}
\pubyear{2015}
\firstpage{122}
\lastpage{144}
\docsubty{FLA}

\makeatletter
\def\p{{(p)}}
\def\simu{\operatorname{sim}}
\makeatother

\begin{document}
\begin{frontmatter}

\title{Reactive point processes: A new approach to predicting power
failures in underground electrical systems\thanksref{T1}}
\runtitle{Reactive point processes}
\thankstext{T1}{Supported in part by Con Edison, the MIT Energy
Initiative Seed Fund and NSF CAREER Grant
IIS-1053407 to C. Rudin.}

\begin{aug}
\author[A]{\fnms{\c{S}eyda}~\snm{Ertekin}\corref{}\ead[label=e1]{seyda@mit.edu}\thanksref{M1}},
\author[A]{\fnms{Cynthia}~\snm{Rudin}\ead[label=e2]{rudin@mit.edu}\thanksref{M1}}
\and
\author[B]{\fnms{Tyler H.}~\snm{McCormick}\ead[label=e3]{tylermc@u.washington.edu}\thanksref{M2}}
\runauthor{\c{S}. Ertekin, C. Rudin and T.~H. McCormick}

\affiliation{Massachusetts Institute of Technology\thanksmark{M1} and
University of Washington\thanksmark{M2}}

\address[A]{\c{S}. Ertekin\\
C. Rudin\\
MIT Computer Science\\
\quad and Artificial Intelligence Laboratory\\
and\\
MIT Sloan School of Management\\
Massachusetts Institute of Technology\\
Cambridge, Massachusetts 02139\\
USA\\
\printead{e1}\\
\phantom{E-mail:\ }\printead*{e2}}

\address[B]{T.~H. McCormick\\
Department of Statistics\\
University of Washington\\
Seattle, Washington 98195\\
USA\\
\printead{e3}}
\end{aug}
%

\received{\smonth{7} \syear{2014}}
\revised{\smonth{9} \syear{2014}}

%
\begin{abstract}
Reactive point processes (RPPs) are a new statistical model designed
for predicting discrete events in time based on past history. RPPs were
developed to handle an important problem within the domain of
electrical grid reliability: short-term prediction of electrical grid
failures (``manhole events''), including outages, fires, explosions and
smoking manholes, which can cause threats to public safety and
reliability of electrical service in cities. RPPs incorporate
self-exciting, self-regulating and saturating components. The
self-excitement occurs as a result of a past event, which causes a
temporary rise in vulner
ability to future events. The self-regulation
occurs as a result of an external inspection which temporarily lowers
vulnerability to future events. RPPs can saturate when too many events
or inspections occur close together, which ensures that the probability
of an event stays within a realistic range. Two of the operational
challenges for power companies are (i) making continuous-time failure
predictions, and (ii) cost/benefit analysis for decision making and
proactive maintenance. RPPs are naturally suited for handling both of
these challenges.
We use the model to predict power-grid failures in Manhattan over a
short-term horizon, and to provide a cost/benefit analysis of different
proactive maintenance programs.
\end{abstract}

%
\end{frontmatter}
\section{Introduction}\label{sec1}
We present a new statistical model for predicting discrete events over
time, called Reactive Point Processes (RPPs). RPPs are a natural fit
for many different domains, and their development was motivated by the
problem of predicting serious events (fires, explosions, power
failures) in the underground electrical grid of New York City (NYC). In
New York City and in other major urban centers, power-grid reliability
is a major source of concern, as demand for electrical power is
expected to soon exceed the amount we are able to deliver with our
current infrastructure [\citet{DOEReport,LisaRhodes,NYBCReport}]. Many
American electrical grids are massive and have been built gradually
since the time of Thomas Edison in the 1880s. For instance, in
Manhattan alone, there are over 21,216 miles of underground cable,
which is almost enough cable to wrap once around the earth. Manhattan's
power distribution system is the oldest in the world, and NYC's power
utility company, Con Edison, has cable databases that started in the
1880s. Within the last decade, in order to handle increasing demands on
NYC's power-grid and increasing threats to public safety, Con Edison
has developed and deployed various proactive programs and policies
[\citet{TheVillager}]. In Manhattan, there are approximately 53,000
access points to the underground electrical grid, which are called
electrical service structures or manholes. Problems in the underground
distribution network are manifested as problems within manholes, such
as underground burnouts or serious events. A multi-year, ongoing
collaboration to predict these events in advance was started in 2007
[\citeauthor{Rudin2014} (\citeyear{RudinETAL2010,RudinEtAl12,Rudin2014})],
where diverse historical
data were used to predict manhole events over a long-term horizon, as
the data were not originally processed enough to predict events in the
short term. Being able to predict manhole events accurately in the
short term could immediately lead to reduced risks to public safety and
increased reliability of electrical service. The data from this
collaboration have sufficiently matured due to iterations of the
knowledge discovery process and maturation of the Con Edison
inspections program, and, in this paper, we show that it is indeed
possible to predict manhole events to some extent within the short term.

The fact that RPPs are a generative model allows them to be used for
cost-benefit analysis, and thus for policy decisions. In particular,
since we can use RPPs to simulate power failures into the future, we
can also simulate various inspection policies that the power company
might implement. This way we can create a robust simulation setup for
evaluating the relative costs of different inspection policies for NYC.
This type of cost-benefit analysis can quantify the cost of the
inspections program as it relates to the forecasted number of manhole events.

RPPs capture several important properties of power failures on the grid:
\begin{itemize}
\item There is an instantaneous rise in vulnerability to future serious
events immediately following an occurrence of a past serious event, and
the vulnerability gradually fades back to the baseline level. This is a
type of \textit{self-exciting} property.
\item There is an instantaneous decrease in vulnerability due to an
inspection, repair or other action taken. The effect of this inspection
fades gradually over time. This is a \textit{self-regulating} property.
\item The cumulative effect of events or inspections can saturate,
ensuring that vulnerability levels never stray too far beyond their
baseline level. This captures \emph{diminishing returns} of many events
or inspections in a row.
\item The baseline level can be altered if there is at least one past event.
\item Vulnerability between similar entities should be similar. RPPs
can be incorporated into a Bayesian framework that shares information
across observably similar entities.
\end{itemize}
RPPs extend self-exciting point processes (SEPPs), which have only the
self-exciting property mentioned above.
Self-exciting processes date back at least to the 1960s [\citet{Bartlett1963,Kerstan1964}]. The applicability of self-exciting point
processes for modeling and analyzing time-series data has stimulated
interest in diverse disciplines, including seismology
[\citeauthor{Ogata1988} (\citeyear{Ogata1988,Ogata1998})],
criminology [\citet
{Porter2012,Mohler2011,Egesdal2010,Lewis2010,Louie2010}], finance [\citet{Chehrazi2011,Ait2010,Bacry2013,Filimonov2012,Embrechts2011,Hardiman2013}],
computational neuroscience [\citet{Johnson1996,Krumin2010}], genome
sequencing [\citet{Reynaud2010}] and social networks [\citet{Crane2008,Mitchell2009,Simma2010uk,Masuda2012,Du2013}]. These models
appear in so many different domains because they are a natural fit for
time-series data where one would like to predict discrete events in
time, and where the occurrence of a past event gives a temporary boost
to the probability of an event in the future. A~recent work on Bayesian
modeling for dependent point processes is that of \citet{GuttorpThor2012}. Paralleling the development of frequentist
literature, many Bayesian approaches are motivated by data on natural
events. \citet{Peruggia1996wn}, for example, develop a Bayesian
framework for the Epidemic-Type-Aftershock-Sequences (ETAS) model.
Nonparametric Bayesian approaches for modeling data from nonhomogeneous
point pattern data have also been developed [see \citet{Taddy2012vj},
e.g.]. \citet{Blundell2012uz} present a nonparametric Bayesian
approach that uses Hawkes models for relational data. An expanded
related work section appears in the supplementary material [\citet{ErtekinSupp}].

The self-regulating property can be thought of as the effect of an
inspection. Inspections are made according to a predetermined policy of
an external source, which may be deterministic or random. In the
application that self-exciting point processes are the most well known
for, namely, earthquake modeling, it is not possible to take an action
to preemptively reduce the risk of an earthquake; however, in other
applications it is clearly possible to do so. In our power failure
application, power companies can perform preemptive inspections and
repairs in order to decrease electrical grid vulnerability. In
neuroscience, it is possible to take an action to temporarily reduce
the firing rate of a neuron. There are many actions that police can
take to temporarily reduce crime in an area (e.g., temporary increased
patrolling or monitoring). In medical applications, doses of medicine
can be preemptively applied to reduce the probability of a cardiac
arrest or other event. Alternatively, for instance, the self-regulation
can come as a result of the patient's lab tests or visits to a physician.

Another way that RPPs expand upon SEPPs is that they allow deviations
from the baseline vulnerability level to saturate. Even if there are
repeated events or inspections in a short period of time, the
vulnerability level still stays within a realistic range. In the
original self-exciting point process model, it is possible for the
self-excitation to escalate to the point where the probability of an
event gets very close to one, which is generally unrealistic. In RPPs,
the saturation function prevents this from happening. Also, if many
inspections are done in a row, the vulnerability level does not drop to
zero, and there are diminishing returns for the later ones because of
the saturation function.

\subsection*{Outline of paper}
We motivate RPPs using the power-grid application in Section~\ref{sec:description_of_data}. We first introduce the general form of the
RPP model in Section~\ref{SecRPPModel}. 
We discuss a Bayesian framework for fitting RPPs in Section~\ref{SecFitting}. The Bayesian formulation, which we implement using
Approximate Bayesian Computation (ABC), allows us to share information
across observably similar entities (manholes in our case). 
For both methods we fit the model to NYC data and performed simulation
studies. Section~\ref{SecPredictionExperiment} contains a prediction
experiment, demonstrating the RPPs' ability to predict future events in
NYC. Once the RPP model is fit to data from the past, it can be used
for simulation. In particular, we can simulate various inspection
policies for the Manhattan grid and examine the costs associated with
each of them in order to choose the best inspection policy. Section~\ref{SecInspectionPolicies} shows this type of simulation using the RPP,
illustrating how it is able to help choose between different inspection
policies, and thus assist with broader policy decisions for the NYC
inspections program. The paper's supplementary material [\citet
{ErtekinSupp}] includes a related work section, conditional frequency
estimator (CF estimator) for the RPP, experiments with a maximum
likelihood approach, a~description of the inspection policy used in
Section~\ref{SecInspectionPolicies} and simulation studies for
validating the fitting techniques for the models in the paper. It also
includes a description and link for a publicly available simulated data
set that we generated, based on statistical properties of the Manhattan
data set.

A short version of this paper appeared in the late-breaking
developments track of AAAI-13 [\citet{Ertekin2013}].

\section{Description of data}
\label{sec:description_of_data}
The data used for the project includes records from the Emergency
Control Systems (ECS) trouble ticket system of Con Edison, which
includes records of responses to past events (total 213,504 records for
53,525 manholes from 1995 until 2010). Part of the trouble ticket for a
manhole fire is in Figure~\ref{Fig:ticket}.
%


\begin{figure}
\fontsize{9}{11}{\selectfont
\begin{verbatim}
FDNY/250 REPORTS F/O 45536 E.51 ST & BEEKMAN PL...MANHOLE FIRE
MALDONADO REPORTS F/O 45536 E.51 ST FOUND SB-9960012 SMOKING
HEAVY...ACTIVE...SOLID...ROUND...NO STRAY VOLTAGE...29-L...
SNOW...FLUSH REQUESTED...ORDERED #100103.
12/22/09 08:10 MALDONADO REPORTS 3 2WAY-2WAY CRABS COPPERED
CUT OUT & REPLACED SAME. ALSO STATES 5 WIRE CROSSING COMES U
P DEAD WILL INVESTIGATE IN SB-9960013.
FLUSH # 100116 ORDERED FOR SAME
12/22/09 14:00 REMARKS BELOW WERE ADDED BY 62355
12/22/09 01:45 MASON REPORTS F/O 4553 E.51ST CLEARED ALL
B/O-S IN SB9960013 ALSO FOUND A MAIN MISSING FROM THE WEST IN
12/22/09 14:08 REMARKS BELOW WERE ADDED BY 62355
SB9960011 F/O 1440 BEEKMAN................................JMC
\end{verbatim}}
\caption{Part of the ECS remarks from a manhole fire ticket in 2009.
The ticket implies that the manhole was actively smoking upon the
worker's arrival. The worker located a crab connector that had melted
(``coppered'') and a cable that was not carrying current (``dead'').
Addresses and manhole numbers were changed for the purpose of
anonymity.\label{Fig:ticket}}
\end{figure}

Events can include serious problems such as manhole fires or
explosions, or nonserious events such as wire burnouts.
These tickets are heavily processed into a structured table, where each
record indicates the time, manhole type (``service box'' or ``manhole,''
and we refer to both types as manholes colloquially), the unique
identifier of the manhole and details about the event. The trouble
tickets are classified automatically as to whether they represent
events (the kind we would like to predict and prevent) or not (in which
case the ticket is irrelevant and removed). The processing of tickets
is based on a study where Con Edison engineers manually labeled
tickets, and is discussed further by \citet{PassonneauEtAlSustKDD11}.\vadjust{\goodbreak}

\begin{figure}
\centering
\begin{tabular}{@{}cc@{}}

\includegraphics{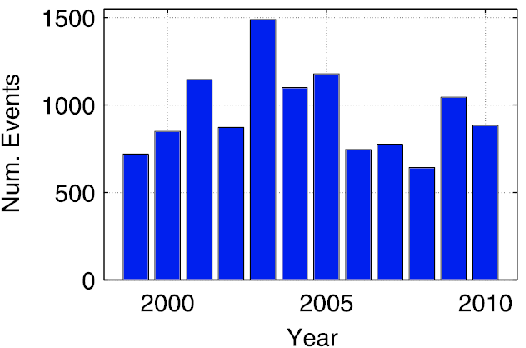}
 & \includegraphics{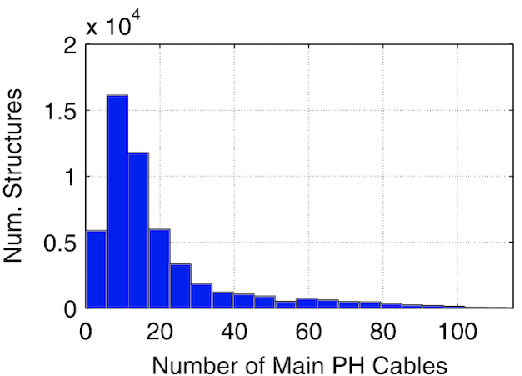}\\
\footnotesize{(a)} & \footnotesize{(b)}
\end{tabular}\vspace*{3pt}
\centering
\begin{tabular}{@{}c@{}}

\includegraphics{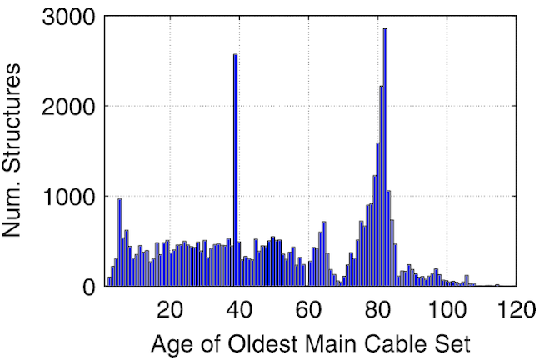}
\\
\footnotesize{(c)}
\end{tabular}
\caption{A plot of the number of yearly events, a histogram of the
number of Main Phase (PH) cables, and a histogram of the age of oldest
cable set in a manhole.}\label{fig:dataset_histograms}
\end{figure}

We have more or less complete event data from 1999 until the present,
and incomplete event data between 1995 and 1999.
A plot of the total number of events per year (using our definition of
what constitutes an event) is provided in Figure~\ref{fig:dataset_histograms}(a).

%
%
%
%
%

We also have manhole location and cable record information, which
contains information about the underground electrical infrastructure.
These two large tables are joined together to determine which cables
enter into which manholes. The inferential join between the two tables
required substantial processing in order to correctly match cables with
manholes. \textit{Main cables} are cables that connect two manholes, as
opposed to \textit{service} or \textit{streetlight cables} which
connect to buildings or streetlights. In our studies on long-term
prediction of power failures, we have found that the number of main
phase cables in a manhole is a relatively useful indicator of whether a
manhole is likely to have an event. Figure~\ref{fig:dataset_histograms}(b)
contains a histogram of the number of main phase cables in a manhole.

The electrical grid was built gradually over the last $\sim$130 years,
and, as a result, manholes often contain cables with a range of
different ages. Figure~\ref{fig:dataset_histograms}(c) contains a
histogram of the age of the oldest main cables in each manhole, as
recorded in the database. Cable age is also used as a feature for our
RPP model. Cable ages range from less than a year old to over 100 years
old; Con Edison started keeping records back in the 1880s during the
time of Thomas Edison. We remark that it is not necessarily true that
the oldest cables are the ones most in need of replacement. Many cables
have been functioning for a century and are still functioning reliably.

We also have data from Con Edison's new inspections program.
Inspections can be scheduled in advance, according to a schedule
determined by a state mandate. This mandate currently requires an
inspection for each structure at least once every 5 years. Con Edison
also performs ``ad hoc'' inspections. These occur when a worker is
inside a manhole for another purpose (e.g., to connect a new
service cable) and chooses to fill in an inspection form. The
inspections are broken down into 5 distinct types, depending on whether
repairs are urgent (Level I) or whether the inspector suggests major
infrastructure repairs (Level IV) that are placed on a waiting list to
be completed. Sometimes when continued work is being performed on a
single manhole, this manhole will have many inspections performed
within a relatively small amount of time---hence our need for
``diminishing returns'' on the influence of an inspection that motivates
the saturation function of the RPP model.

Some questions of interest to power companies are as follows:
\begin{longlist}[(ii)]
\item[(i)] Can we predict failures continuously in time, and can we model
how quickly the influence of past events and inspections fade over time?
\item[(ii)] Can we develop a cost/benefit analysis for proactive maintenance policies?
\end{longlist}
RPPs will help with both of these questions.

%

\section{The reactive point process model}\label{SecRPPModel}
We begin with a simpler version of RPPs where there is only one
time-series corresponding to a single entity (manhole). Our data
consist of a series of $N_E$ events with event times $t_{1}, t_{2},\ldots,
t_{N_E}$ and a series of given inspection times denoted by $\bar
{t}_{1}, \bar{t}_{2},\ldots, \bar{t}_{N_I}$. The inspection times are
assumed to be under the control of the experimenter. RPPs model events
as being generated from a nonhomogeneous Poisson process with intensity
$\lambda(t)$ where
%
\begin{equation}\quad
\label{RPPmodel} \lambda(t)=\lambda_{0} \biggl[1+g_{1}
\biggl( \sum_{\forall
t_{e}<t}g_{2}(t-t_{e})
\biggr)-g_{3} \biggl(\sum_{\forall\bar
{t}_{i}<t}g_{4}(t-
\bar{t}_{i}) \biggr)+C_1{\mathbf1}_{[N_E\geq1]}
\biggr],
\end{equation}
where $t_{e}$ are event times and $\bar{t}_{i}$ are inspection times.
The vulnerability level permanently goes up by $C_1$ if there is at
least one past event, where $C_1$ is a constant that can be fitted. The
$C_1{\mathbf 1}_{[N_E\geq1]}$ term is present to deal with ``zero
inflation,'' where the case of zero events needs to be handled
separately than one or more past events. Functions $g_2$ and $g_4$ are
the self-excitation and self-regulation functions, which have initially
large amplitudes and decay over time. Self-exciting point processes
have only $g_2$, and not the other functions, which are novel to RPPs.
Functions $g_1$ and $g_3$ are the saturation functions, which start out
as the identity function and then flatten farther from the origin. If
the total sum of the excitation terms is large, $g_1$ will prevent the
vulnerability level from increasing too much. Similarly, $g_4$ controls
the total possible amount of self-regulation and encodes ``diminishing
returns'' for having several inspections in a row.

\begin{figure}
\centering
\begin{tabular}{@{}cc@{}}

\includegraphics{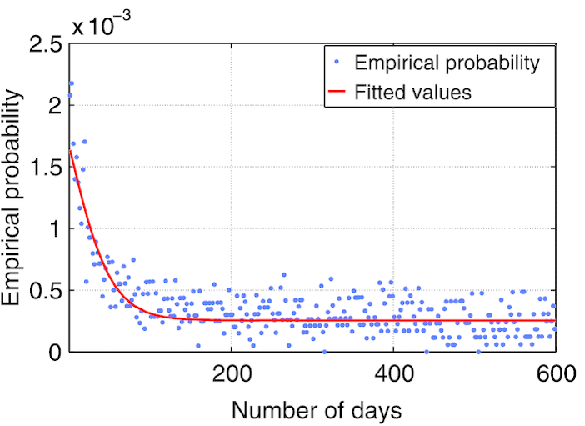}
 & \includegraphics{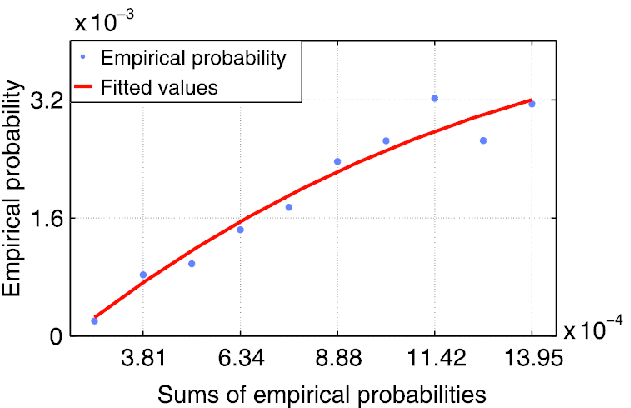}\\
\footnotesize{(a) Empirical probabilities and fitted values} & \footnotesize{(b) Empirical probabilities and fitted values}\\
\footnotesize{for the
self-excitation function $g_2$} & \footnotesize{for the saturation
function $g_1$}
\end{tabular}
\caption{Fitted functions for empirical probabilities for the Manhattan
data set. These figures display results for the conditional frequency
estimator used to derive the form of the RPP model. The left figure
shows the empirical probability of another event given a previous event
a given number of days in the past. The decreasing empirical
probability with time motivates our self-excitation function. The right
plot shows the increase in propensity for another event given the total
cumulative probability from past events. The curvature indicates that
additional events have diminishing returns on the likelihood of another
event, motivating the saturation component of the RPP.}
\label{fig:fitted_g1_g2_realdata}
\end{figure}

\begin{figure}
\centering
\begin{tabular}{@{}cc@{}}

\includegraphics{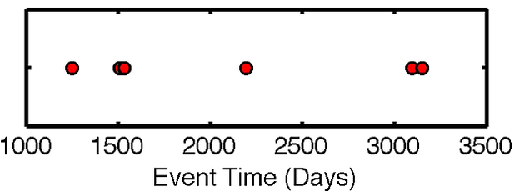}
 & \includegraphics{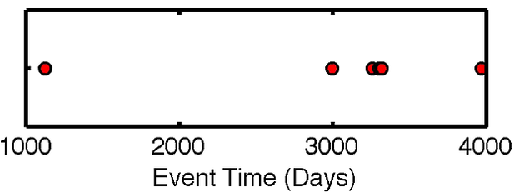}\\[3pt]

\includegraphics{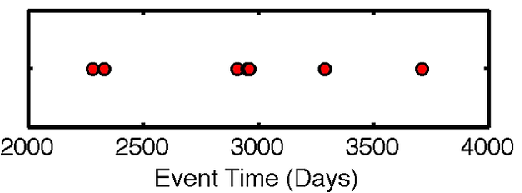}
 & \includegraphics{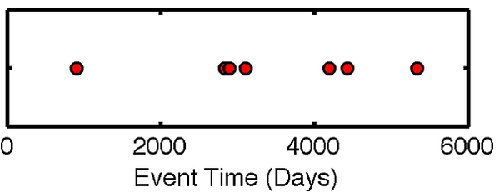}\\
\end{tabular}
\caption{Time of events in distinct manholes in the Manhattan data that
demonstrate the self-excitating behavior. The $x$-axis is the number of
days elapsed since the day of first record in the data set and the
markers indicate the actual time of events.}
\label{fig:manhattan_timeline}
\end{figure}

The RPP model arose based on exploratory work performed using a
conditional frequency (CF) estimator of the data. To construct the CF
estimator, we computed the empirical probability of another event
occurring on a day $t$ given that an event occurred at $t=0$. To obtain
these probabilities, we first align the sequences of time so that $t=0$
represents the time when an event happened. We now have a series of
``trails'' that give the probability of another event, conditional on
the last event that occurred for a given manhole. We used only trails
that were far apart in time so we could look at the effect of each
event without considering short-term influences of other previous
events. What we see from Figure~\ref{fig:fitted_g1_g2_realdata}(a) is that
the conditional probability for experiencing a second event soon after
the first event is high and decays with $t$. This decay represents
self-exciting behavior. To see evidence of self-excitation from the raw
data, we present plots of event times for several manholes in Figure~\ref{fig:manhattan_timeline}. We see a clear grouping of events which
is consistent with self-exciting behavior.
These observations lead us to include the $g_2$ term in~(\ref
{RPPmodel}). The behavior we observe could not be easily explained
using a simple random effects model; an attempt to do this is within
Section~4 of the supplementary material [\citet
{ErtekinSupp}].

%

Next, we evaluate whether subsequent events continue to increase
propensity for another event or whether the risk in the most troubled
manholes ``saturates'' and multiple manhole events in a row have
diminishing-returns on the conditional probabilities.
Figure~\ref{fig:fitted_g1_g2_realdata}(b) shows the saturation effect. The
$y$-axis of this plot contains raw empirical probabilities of another
event. The $x$-axis are sums of effects from previous recent events (sums
of $g_2$ values).
If Figure~\ref{fig:fitted_g1_g2_realdata}(b) were linear, we would not see
diminishing returns. That is, a linear trend in
Figure~\ref{fig:fitted_g1_g2_realdata}(b) would indicate that each subsequent event
increases the likelihood of another event by the same amount. Instead,
we see a distinct curve, indicating that the additional increase in
risk decreases as the number of events rises. To further aid in
developing a functional form of the model, we fit smooth curves to the
data displayed in Figure~\ref{fig:fitted_g1_g2_realdata}(a) and (b).
The process for fitting these smooth curves,
as well as simulation experiments for validation, is described in
detail in the supplementary material [\citet{ErtekinSupp}]. The fitted
values for the smooth curves are
\begin{eqnarray*}
g_2(t)&=& \frac{11.62}{1 + e^{0.039t}},
\\
g_1(t) &=& 16.98\times \biggl(1-\log \bigl(1+e^{-0.15t} \bigr)
\times \frac
{1}{\log2} \biggr).
\end{eqnarray*}
These estimates inspired the parameterizations we provided in equation
(\ref{paraform}). We also estimated the baseline hazard rate $\lambda
_0$ and baseline change $C_1$ for Manhattan as $\lambda_0 =
2.4225\times10^{-4}$ and $C_1= 0.0512$.

Because the inspection program is relatively new, we were not able to
trace out the full functions $g_4$ and $g_3$; however, we strongly
hypothesize that the inspections have an effect that wears off over
time based on a matched pairs study [see, e.g., \citet
{PassonneauEtAlSustKDD11}], where we showed that for manholes that had
been inspected at least twice, the second manhole inspection does not
lead to the same reduction in vulnerability as the first manhole
inspection does. In what follows, we will show how the parameters of
$g_1$, $g_2$, $g_3$ and $g_4$ can be made to specialize to each
individual manhole adaptively.

Inspired by the CF estimator, we use the family of functions below for
fitting power-grid data, where $a_{1}$, $b_{1}$, $a_{3}$, $b_{3}$,
$\beta$ and $\gamma$ are parameters that can be either modeled or
fitted: 
\begin{eqnarray}
\label{paraform}
g_1(\omega) & =& a_{1} \times
\biggl(1 - \frac{1}{\log 2}\log \bigl(1+e^{-b_{1}\omega} \bigr) \biggr),\qquad
g_2(t) = \frac{1}{1+e^{\beta t}},
\nonumber
\\[-8pt]
\\[-8pt]
\nonumber
g_3(\omega) & = & a_{3} \times \biggl(1 -
\frac{1}{\log 2}\log \bigl(1+e^{b_{3}\omega} \bigr) \biggr), \qquad g_4(t)
= \frac{-1}{1+e^{\gamma t}}.
\end{eqnarray}
The factors of $\log 2$ ensure that the vulnerability level is not negative.

We need some notation in order to encode the possibility of multiple
manholes. In the case that there are multiple entities, there are $P$
time-series, each corresponding to a unique entity $p$. For medical
applications, each $p$ is a patient, and for the electrical grid
reliability application, $p$ is a manhole. Our data consist of events
$\{t\p_{e}\}_{p,e}$, inspections $\{\bar{t}\p_i\}_{p,i}$ and,
additionally, we may have covariate information $M_{p,j}$ about every
entity $p$, with covariates indexed by $j$. Covariates for the medical
application might include a patient's gender, age at the initial time,
race, etc. For the manhole events application, covariates include the
number of main phase cables in the manhole (number of current carrying
cables between two manholes), the total number of cable sets (total
number of bundles of cables) including main, service and streetlight
cables, and the age of the oldest cable set within the manhole. All
covariates were normalized to be between $-0.5$ and 0.5.

Within the Bayesian framework, we can naturally incorporate the
covariates to model functions $\lambda_p$ for each $p$ adaptively.
Consider $\beta$ in the expression for the self-excitation function
$g_2$ above. The $\beta$ terms depend on individual-level covariates.
In notation,
%
\begin{equation}
\label{gbetagamma} g_2^{(p)}(t) = \frac{1}{1+e^{\beta^{(p)} t}},\qquad
g_4^{(p)}(t) = \frac{-1}{1+e^{\gamma^{(p)} t}}.
\end{equation}
The $\beta^{(p)}$'s are assumed to be generated via a hierarchical
model of the form
\[
\bolds{\beta}=\log \bigl(1+e^{-\mathbf M\bolds{\upsilon}} \bigr)\qquad\mbox{where }
\bolds{\upsilon}\sim N \bigl(0,\sigma^{2}_{\upsilon} \bigr)
\]
are the regression coefficients and $\mathbf{M}$ is the matrix of observed
covariates. The $\gamma^{(p)}$'s are modeled hierarchically in the same
manner,
\[
\bolds{\gamma}=\log \bigl(1+e^{-\mathbf M\bolds{\omega}} \bigr)\qquad\mbox{with }
\bolds {\omega}
\sim N \bigl(0,\sigma^2_{\omega} \bigr).
\]
This permits slower or faster decay of the self-exciting and
self-regulating components based on the characteristics of the
individual. For the electrical reliability application, we have noticed
that manholes with more cables and older cables tend to have faster
decay of the self-exciting terms, for instance.

\subsection*{Demonstrating the need for the saturation function in the
RPP model}
\label{demonstrating_the_need}
In the previous section we used exploratory tools on the Manhattan data
to demonstrate diminishing returns in risk for multiple subsequent
events. In what follows, we link the exploratory work in the last
section with our modeling framework, demonstrating how the standard
linear self-exciting process can produce unrealistic results under
ordinary conditions.

\begin{figure}[b]
\centering
\begin{tabular}{@{}c@{\quad}c@{}}

\includegraphics{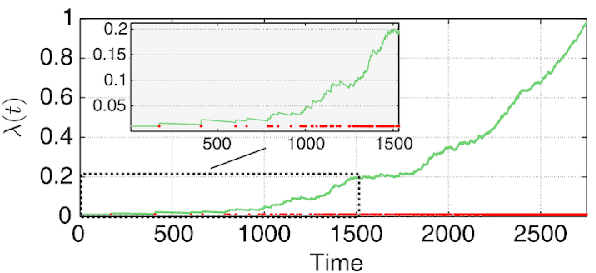}
 & \includegraphics{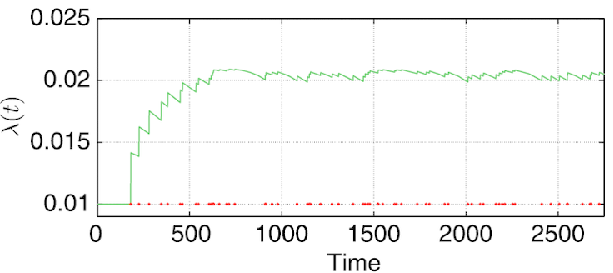}\\
\footnotesize{(a) Model with self-excitation function $g_2$} & \footnotesize{(b) Model with self-excitation function $g_2$}\\
\footnotesize{(without saturation function $g_1$ and inspections)} & \footnotesize{and saturation function $g_1$ (no inspections)}\\
\footnotesize{and a zoomed view of the first 1500 days} & \footnotesize{}\\[5pt]

\includegraphics{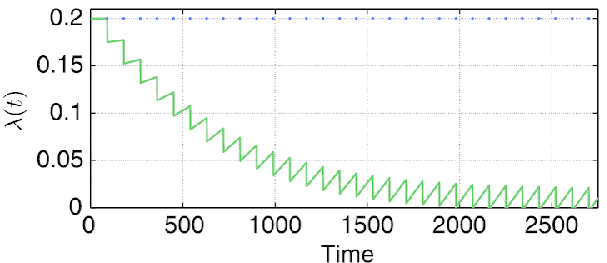}
 & \includegraphics{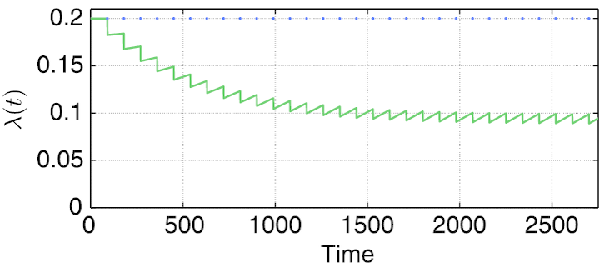}\\
\footnotesize{(c) Model with self-regulation function $g_4$} & \footnotesize{(d) Model with self-regulation function $g_4$}\\
\footnotesize{(without saturation function $g_3$ and events)} &
\footnotesize{and saturation function $g_3$ (no events)}
\end{tabular}
\caption{The effect of the saturation functions $g_1$ and $g_3$. The
dots on the time axis in subfigures \textup{(a)} and
\textup{(b)} indicate the times of events, and the dots in
subfigures \textup{(c)} and \textup{(d)}
indicate the times of inspections. The figures on the right include
saturation, and the figures on the left do not include saturation.
Without saturation, the self-excitation function in \textup{(a)} grows unbounded,
whereas the self-regulation
function in \textup{(c)} drops to an unrealistic level of
zero. The effects of the saturation functions in Figures \textup{(b)} and
\textup{(d)} keep $g_2$ and $g_4$
within realistic bounds.}
\label{fig:simulated_g1_g2_g3_g4}
\end{figure}

First we show that the self-excitation term can cause the rate of
events $\lambda(t)$ to increase without bound. To show this, we
considered a baseline vulnerability of $\lambda_0=0.01$, setting
$C_1=0.1$, used $g_2(t) = \frac{1}{1+e^{0.005t}}$, and omitted the
other components of the model (no inspections, no saturation $g_1$).
The self-excitation eventually causes the rate of events to escalate
unrealistically as shown in Figure~\ref{fig:simulated_g1_g2_g3_g4}
(upper left). The embedded subfigure is a zoomed-in version of the
first 1500 time steps.

When we include the saturation function $g_1$, the excitation is
controlled, and the probability of an event no longer increases to
unreasonable levels. We used $g_1(\omega) = 1- \frac{1}{\log2}\log
(1+e^{-\omega})$, so that the vulnerability $\lambda(t)$ can reach to a
maximum value of 0.021. The result is in Figure~\ref{fig:simulated_g1_g2_g3_g4} (upper right).

Now we show the effect of the saturation function $g_3$ in the presence
of repeated inspections. If no manhole events occur and the manhole is
repeatedly inspected, then using the linear SEPP model, its
vulnerability levels can become arbitrarily close to 0. This is not
difficult to show, and we do this in Figure~\ref{fig:simulated_g1_g2_g3_g4} (lower left). Here we used $\lambda_0 =
0.2$, $g_4(t) = \frac{-0.25}{1+e^{0.002t}}$, and omitted $g_3$. We ran
the same experiment but with saturation, specifically, with $g_3(\omega
) = 1- \frac{1}{\log2}\log(1+e^{\omega})$. The results in Figure~\ref{fig:simulated_g1_g2_g3_g4} (lower right) show that the saturation
function never lets the vulnerability drop unrealistically far below
the baseline level.

%
%
%
%
%
%

\section{Fitting RPP statistical models}\label{SecFitting}
In this section we describe our Bayesian framework for inference using
RPP models. The RPP intensity in equation~(\ref{RPPmodel}) provides
structure to capture self-excitation, self-regulation and saturation.
First, in Section~\ref{SecLikelihood} we describe the likelihood for
the RPP statistical model. We then describe prior distributions and our
computational strategy for sampling from the posterior in Section~\ref{SecABC}. Section~\ref{RPPcompare} then details the values we use in
making predictions. Along with the results presented here, we
extensively evaluated our inference strategy using a series of
simulation experiments, where the goal is to recover parameters of
simulated data for which there is ground truth. We further applied the
method of maximum likelihood to the Manhattan power-grid data. Details
of these additional experiments are in the supplementary material [\citet
{ErtekinSupp}].

\subsection{RPP likelihood}\label{SecLikelihood}
This section describes the likelihood for the RPP statistical model.
Using the intensity function described in Section~\ref{SecRPPModel},
the RPP likelihood is derived using the likelihood formula for a
nonhomogeneous Poisson process over the time interval $[0,T_{\max}]$:
\begin{eqnarray}\label{eq:likwithcovars}
&&\log\mathcal{L} \bigl( \bigl\{t_1^{\p},\ldots,
t_{N_E^{\p}}^{\p
} \bigr\}_p; \bolds{\upsilon},
a_{1}, \mathbf{M} \bigr)
\nonumber
\\[-8pt]
\\[-8pt]
\nonumber
&&\qquad=\sum_{p=1}^P
\Biggl[\sum_{e=1}^{{N_E^{\p}}}\log \bigl(
\lambda_{p} \bigl(t_e^{\p} \bigr) \bigr)-\int
_{0}^{T_{\max
}}\lambda_{p}(u)\,du
\Biggr],
\end{eqnarray}
where $\bolds\upsilon$ are coefficients for covariates represented by the
matrix $\mathbf{M}$. The covariates are the number of main phase cables in
the manhole (number of current carrying cables between two manholes),
the total number of cable sets (total number of bundles of cables)
including main, service and streetlight cables, and the age of the
oldest cable set within the manhole. All covariates were normalized to
be between $-0.5$ and 0.5.

\subsection{Bayesian RPP}\label{SecABC}
Developing a Bayesian framework facilitates sharing of information
between observably similar manholes, thus making more efficient use of
available covariate information. The RPP model encodes much of our
prior information into the shape of the rate function given in
equation~(\ref{RPPmodel}). As discussed in Section~\ref{SecRPPModel}, we
opted for a simple, parsimonious model that imposes mild regularization
and information sharing without adding substantial additional
information; specifically, we use diffuse Gaussian priors on the log
scale for each regression coefficient.

We fit the model using Approximate Bayesian Computation~[\citet
{Diggle1984vf}]. The principle of Approximate Bayesian Computation
(ABC) is to randomly choose proposed parameter values, use those values
to generate data, and then compare the generated data to the observed
data. If the difference is sufficiently small, then we accept the
proposed parameters as draws from the approximate posterior. To do ABC,
we need two things: (i) to be able to simulate from the model and (ii)
a summary statistic. To compare the generated and observed data, the
summary statistic from the observed data, $S(\{t_1^{\p},\ldots, t_{N_E^{\p
}}^{\p}\}_p)$, is compared to that of the data simulated from the
proposed parameter values, $S(\{t_1^{\p,\simu},\ldots,t_{N_E^{\p,\simu
}}^{\p,\simu}\}_p)$. If the values are similar, it indicates that the
proposed parameter values may yield a useful model for the data.

A critical difference between updating a parameter value in an ABC
iteration versus, for example, a Metropolis--Hastings step is that ABC
requires simulating from the likelihood, whereas Metropolis--Hastings
requires evaluating the likelihood. In our context, we are able to both
evaluate and simulate from the likelihood with approximately the same
computational complexity. ABC has some advantages, namely, that we have
meaningful summary statistics, discussed below. Further, in our case it
is not particularly computationally challenging, as
we already extensively simulate from the model as a means of evaluating
hypothetical inspection policies.
We evaluated the adequacy of this method extensively in simulation
studies presented in the supplementary material [\citet{ErtekinSupp}].

A key conceptual aspect of ABC is that one can choose the summary
statistic to best match the problem. The sufficient statistic for the
RPP is the vector of event times, and thus gives no data reduction---so
we choose other statistics. One important insight in constructing our
summary statistic is that changing the parameters in the RPP model
alters the distribution of times between events. The histogram of time
differences for a homogenous Poisson Process, for example, has an
exponential decay. The self-exciting process, on the other hand, has a
distribution resembling a lognormal because of the positive association
between intensities after an event occurs. Altering the parameters of
the RPP model changes the intensity of self-excitation and
self-regulation, thus altering the distribution of times between
events. We construct our first statistic, therefore, by examining the
KL divergence between the distribution of times between events in the
data and the distribution between event times in the simulated data. We
do this for each of our proposed parameters. Examining the distribution
of times between events, though not the true sufficient statistic,
captures a concise and low-dimensional summary of a key feature of the
process. This statistic does not, however, capture the overall
prevalence of events in the process. Since we focus only on the \emph
{distribution} of times between events, various processes with
different overall intensity could produce distributions with similar KL
divergence to the data distribution. We therefore introduce a second
statistic that counts the total number of events. We contend that
together these statistics represent both the spacing and the overall
scale (or frequency) of events. Thus, the two summary measures we use
are as follows:
\begin{longlist}[1.]
\item[1.] DNE: The difference in the number of events in the simulated and
observed data.
\item[2.] KL: The Kullback--Leibler divergence between two histograms, one
from the observed data and one from the real data. These are histograms
of time differences between events.
\end{longlist}


\begin{figure}[b]
\centering
\begin{tabular}{@{}cc@{}}

\includegraphics{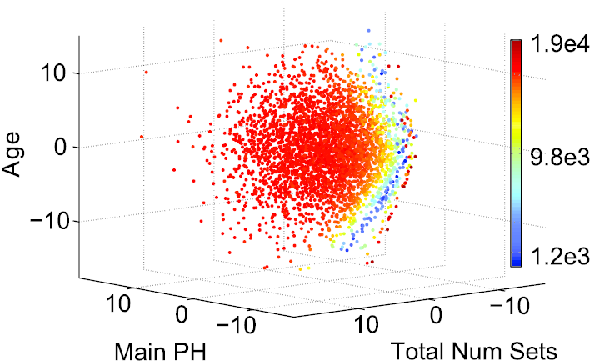}
 & \includegraphics{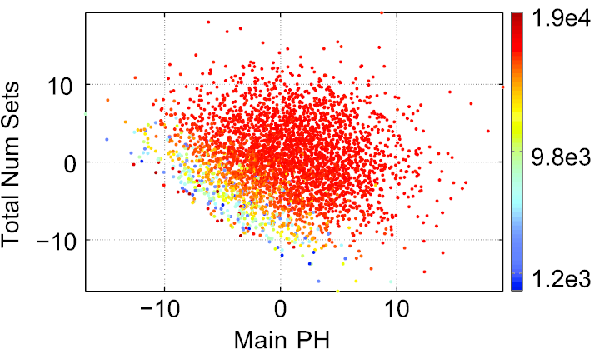}\\
\end{tabular}
\caption{DNE for Manhattan data set. Each axis corresponds to the
coefficient for one of the covariates. The magnitude of DNE is
indicated by the color.}
\label{Figure:DNE3D2D}
\end{figure}

\begin{figure}
\centering
\begin{tabular}{@{}cc@{}}

\includegraphics{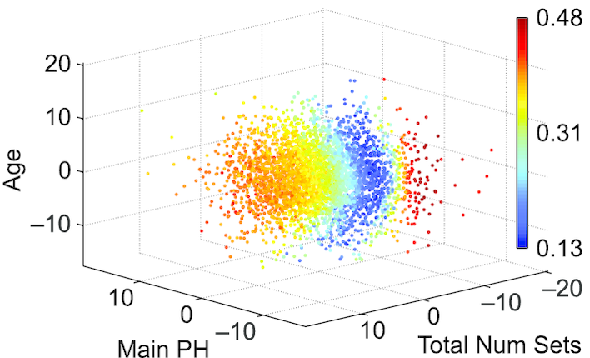}
 & \includegraphics{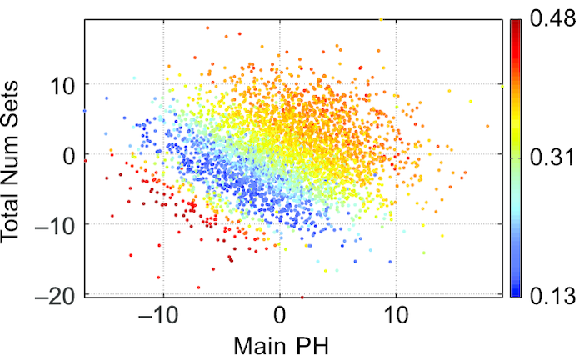}\\
\end{tabular}
\caption{KL for Manhattan data set. Each axis corresponds to the
coefficient for one of the covariates. The magnitude of KL is indicated
by the color.}
\label{Figure:KL3D2D}
\end{figure}

For the NYC data, we visualized three-dimensional parameter values,
both for DNE (in Figure~\ref{Figure:DNE3D2D}) and KL (in Figure~\ref{Figure:KL3D2D}) metrics. In both figures, smaller values (dark blue)
are better. As seen, the regions where KL and DNE are optimized are
very similar.

Denoting the probability distribution of the actual data as $P$ and the
probability distribution of the simulated data as $Q_{\bolds\upsilon}$, KL
Divergence is computed as
\[
{\mathrm{KL}}(P\Vert Q_{\bolds\upsilon}) = \sum_{\mathrm{bin}}
\ln \biggl( \frac{P(\mathrm{bin})}{Q_{\bolds\upsilon}(\mathrm
{bin})} \biggr)P(\mathrm{bin}).
\]

As mentioned previously in this section, the Bayesian portion of our
model is relatively parsimonious but does impose mild regularization
and encourages stability. We require a distribution $\pi$ over
parameter values. If covariates are not used, $\pi$ is a distribution
over $\beta$ (and $\gamma$ if inspections are present). If covariates
are used, $\pi$ is a distribution over $\bolds{\upsilon}$ and $\bolds{\omega
}$. One option for $\pi$ is a uniform distribution across a grid of
reasonable values. Another option, which was used in our experiments,
is to simulate from diffuse Gaussian/weakly informative priors on the
log scale [e.g., draw $\log(\nu_{j})\sim N(0,5)$]. We assumed
that $C_1$ and $a_{1}$ can be treated as tuning constants to be
estimated using the CF estimator method, though it is possible to
define priors on these quantities as well if desired.

%

There is an increasingly large literature in both the theory and
implementation of ABC [see, e.g., \citet
{Fearnhead2012,Beaumont2009,Drovandi2011}] that could be used to
produce estimates of the full posterior. In the supplementary material
[\citet{ErtekinSupp}], we present an importance sampling algorithm as
one possible approach. In our work, however, the goal is to estimate
the posterior mode, which we then use for prediction. To verify our ABC
procedure, we used simulated ground truth data with known $\beta$ and
$\gamma$ values, and attempted to recover these values with the ABC
method, for both the DNE and KL metrics. We performed extensive
simulation studies to evaluate this method and full results are given
in the supplementary material [\citet{ErtekinSupp}].

In the next section we discuss how we estimate the posterior mode by
using a manifold approximation to the region of high posterior density.
We begin by generating a set of proposed parameter values using the
prior distributions. Consistent with ABC, we simulate data from each
set of candidate values and compare the simulated data to our observed
data using the KL and DNE statistics described above. (From here, we
could, e.g., define a kernel and accept draws with a given
probability as in importance sampling. Instead, our goal is estimating
the posterior mode to find parameters for the policy decision, as we
describe next.)


\subsection{Choosing parameter values for the policy decision}\label{RPPcompare}
\begin{figure}

\includegraphics{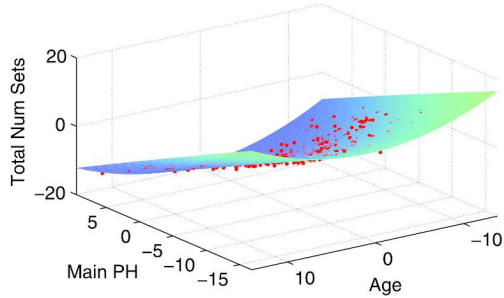}

\caption{Fitted manifold of $\bolds{\upsilon}$ values with smallest KL
divergence and smallest DNE.} \label{fig:kl_fit_poly}
\end{figure}

For the policy simulation in Section~\ref{SecInspectionPolicies} we
wish to choose one set of parameter values to inform our decision. 
In order to choose a single best value of the parameters, we fit a
polynomial manifold to the intersection of the bottom 10\% of KL values
and the bottom 10\% of DNE values.
Defining $\upsilon_1$, $\upsilon_2$ and $\upsilon_3$ as the
coefficients for number of main phase cables, age of oldest main cable
set and total number of sets features, the formula for the manifold is
\begin{eqnarray*}
\upsilon_3 &=& -9.6 -0.98\upsilon_1 -0.13
\upsilon_2 -1.1\times 10^{-3}(\upsilon_1)^2
-3.6\times10^{-3}\upsilon_1\upsilon_2\\
&&{} + 4.67
\times 10^{-2}(\upsilon_2)^2,
\end{eqnarray*}
which is determined by a least squares fit to the data. The fitted
manifold is shown in Figure~\ref{fig:kl_fit_poly} along with the data.

We then optimized for the point on the manifold closest to the origin.
This implicitly adds regularization, as it chooses the parameter values
closest to the origin. This point is $\upsilon_1=-4.6554$, $\upsilon
_2=-0.5716$, and $\upsilon_3=-4.8028$.

Note that cable age (corresponding to the second coefficient) is not
the most important feature defining the manifold. As previous studies
have shown [\citet{RudinETAL2010}], even though there are very old
cables in the city, the age of cables within a manhole is not alone the
best predictor of vulnerability. Now we also know that it is not the
best predictor of the rate of decay of vulnerability back to baseline
levels. This supports Con Edison's goal to prioritize the most
vulnerable components of the power-grid, rather than simply replacing
the oldest components. The features that mainly determine decay of the
self-excitation function $g_2$ are the number of main phase cables and
the number of cable sets. As either or both of these numbers increase,
decay rate $\beta$ increases, meaning that manholes with more cables
tend to return to baseline levels faster than manholes with fewer cables.

\section{Predicting events on the NYC power-grid}\label{SecPredictionExperiment}

\begin{figure}

\includegraphics{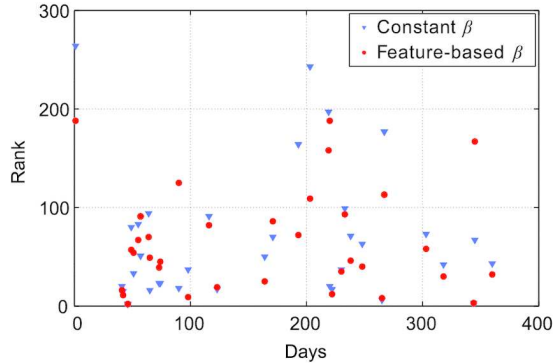}

\caption{Ranking differences between feature-based and constant
(nonfeature-based) $\beta$ strategies.}
\label{fig:adaptive_constant_beta_rankdiff}
\end{figure}

Our first experiment aims to evaluate whether the CF estimator or the
feature-based strategy introduced above is better in terms of
identifying the most vulnerable manholes. To do this, we selected 5000
manholes (rank 1001--6000 from the project's current long-term
prediction model). These manholes have similar vulnerability levels,
which allows us to isolate the self-exciting effect without modeling
the baseline level. Using both the feature-based $\beta$ (ABC, with KL
metric) and constant $\beta$ (CF estimator method) strategies, the
models were trained on data through 2009, and then we estimated the
vulnerabilities of the manholes on December 31st, 2009. These
vulnerabilities were used as the \textit{initial} vulnerabilities for
an evaluation on the 2010 event data. 2010 is a relevant year because
the first inspection cycle ended in 2009. All manholes had been
inspected at least once, and many were inspected toward the end of
2009, which stabilizes the inspection effects. For each of the 53K
manholes and at each of the 365 days of 2010,
when we observed a serious event in a manhole~$p$, we evaluated the
rank of that manhole with respect to both the feature-based and
nonfeature-based models, where rank represents the number of manholes
that were given higher vulnerabilities than manhole $p$.
As our goal is to compare the relative rankings provided by the two
strategies, we consider only events where the vulnerabilities assigned
by both strategies are different than the baseline vulnerability.
Figure~\ref{fig:adaptive_constant_beta_rankdiff} displays the ranks of
the manholes on the day of their serious event. A~smaller rank
indicates being higher up the list, thus lower is better. Overall, we
find that the feature-based $\beta$ strategy performs better than the
nonfeature-based strategy over all of the
rank comparisons in 2010
($p$-value 0.09, sign test). Our results mainly illustrate that using
different decay rates on past events for different types of manholes
leads to better predictions. Recall from Section~\ref{RPPcompare} that
larger manholes tend to recover faster from previous events. The
approach without the features ignores the differences between manholes,
and uses the same decay rate, whereas the feature-based RPP takes these
decay rates into account in making predictions.
%

In the second experiment, we compared the feature-based $\beta$
strategy to the Cox proportional--hazard model, which is commonly used
in survival analysis to assess the probability of failure in mechanical systems.
We employed this model to assess the likelihood of a manhole having a
serious event on a particular day. For each manhole, we used the same
three static covariates as in the feature-based $\beta$ model, and
developed four time-dependent features. The time-varying features for
day $t$ are (1) the number of times the manhole was a trouble hole
(source of the problem) for a serious event until $t$, (2) the number
of times the manhole was a trouble hole for a serious event in the last
year, (3) the number of times the manhole was a trouble hole for a
precursor event (less serious event) until $t$, and (4) the number of
times the manhole was a trouble hole for a precursor event in the last
year. The feature-based $\beta$ model currently does not differentiate
serious and precursor events, though it is a direct extension to do
this if desired.
The model was trained using the coxph function in the R survival
package using data prior to 2009, and then predictions were made on the
test set of 5000 manholes in the 2010 data set. These predictions were
transformed into ranked lists of manholes for each day. We then
compared the ranks achieved by the Cox model with the ranks of manholes
at the time of events. The difference of aggregate ranks was 
in favor of the feature-based $\beta$ approach ($p$-value 7e--06, sign
test), indicating that the feature-based $\beta$ strategy provides a
substantial advantage in its ability to prioritize vulnerable manholes.

The Cox model we compared with represents a ``long-term'' model similar
to what we were using previously for manhole event prediction on Con
Edison data [\citet{RudinETAL2010}]. The Cox model considers different
information, namely, counts of past events. These counts are
time-varying, but the past events do not smoothly wear off in time as
they do for the RPP. The fact that the RPP model is competitive with
the Cox model indicates that the effects of past manhole events do wear
off with time (in agreement with Figure~\ref{fig:fitted_g1_g2_realdata}
where we traced the decay directly using data). The saturating elements
of the model ensure that the model is physically plausible, since we
showed in Section~\ref{demonstrating_the_need} that the results could
be unphysical (with rates going to 0 or above 1) without the saturation.

\section{Making broader policy decisions using RPPs} \label
{SecInspectionPolicies}
Because the RPP model is a generative model, it can be used to simulate
the future, and thus assist with broader policy decisions regarding how
often inspections should be performed. This can be used to justify
allocation of spending. Con Edison's existing inspection policy is a
combination of targeted periodic inspections and ad hoc inspections.
The targeted inspections are planned in advance, whereas the ad hoc
inspections are unscheduled. An ad hoc inspection could be performed
while a utility worker is in the process of, for instance, installing a
new service cable to a building or repairing an outage. Either source
of inspection can result in an urgent repair (Type~I), an important but
not urgent repair (Type II), a suggested structural repair (Types~III
and IV), or no repair, or any combination of repairs. Urgent repairs
need to be completed before the inspector leaves the manhole, whereas
Type IV repairs are placed on a waiting list. According to the current
inspections policy, each manhole undergoes a targeted inspection every
5 years. The choice of inspection policy to simulate can be determined
very flexibly, and any inspection policy and hypothesized effect of
inspections can be examined through simulation.

As a demonstration, we conducted a simulation over a 20 year future
time horizon that permits a cost-benefit analysis of the inspection
program, when targeted inspections are performed at a given frequency.
To do this simulation, we require the following:
\begin{itemize}
\item A characterization of manhole vulnerability. For Manhattan, this
is learned from the past using the ABC RPP feature-based $\beta$
training strategy for the saturation function $g_1$ and the
self-excitation function $g_2$ discussed above. Saturation and
self-regulation functions $g_3$ and $g_4$ for the inspection program
cannot yet be learned due to the newness of the inspection program and
are discussed below.
%
\item An inspection policy. The policy can include targeted, ad hoc or
history-based inspections. We chose to evaluate ``bright line''
inspection policies, where each manhole is inspected once in each $Y$
year period, where $Y$ is varied (discussed below). We also included an
ad hoc inspection policy that visits 3 manholes per day on average.
\end{itemize}

\begin{figure}
\centering
\begin{tabular}{@{}cc@{}}

\includegraphics{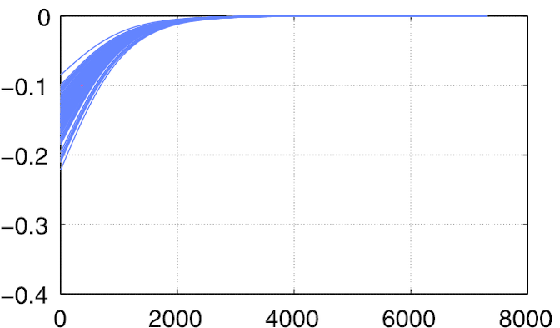}
 & \includegraphics{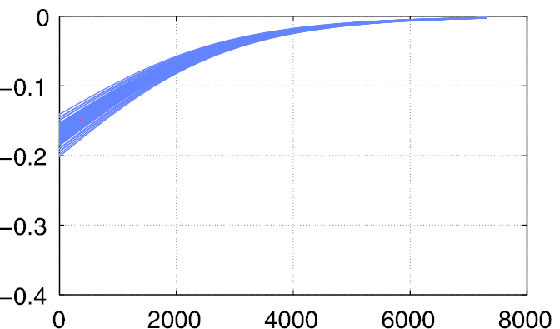}\\
\footnotesize{(a) $g_4$ for Type I inspections} & \footnotesize{(b) $g_4$ for Types II, III, IV}\\
\end{tabular}
\centering
\begin{tabular}{@{}c@{}}

\includegraphics{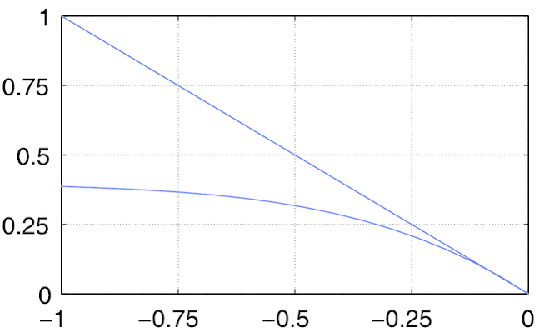}
\\
\footnotesize{(c) $g_3=0.4(1-\log(1+e^{3.75x})\frac{1}{\log 2})$}
\end{tabular}
\caption{Saturation and self-regulation functions $g_3$ and $g_4$ for
simulation.}\label{g3g4simulation}
\end{figure}

\textit{Effect of inspections}: The effect of inspections on
the overall vulnerability of manholes were designed in consultation
with domain experts. The choices are somewhat conservative, so as to
give a lower bound for costs. The effect of an urgent repair (Type I)
is different from the effect of less urgent repairs (Types II, III and
IV). For all inspection types, after 1 year beyond the time of the
inspection, the effect of the inspection decays to, on average, 85\% of
its initial effect, in agreement with a short-term empirical study on
inspections.
(There is some uncertainty in this initial effect, and the initial drop
in vulnerability is chosen from a normal distribution so that after one
year the effect decays to a mean of 85\%.) For Type I inspections, the
effect of the inspection decays to baseline levels after approximately
3000 days, and for Types II, III and IV, which are more extensive
repairs, the effect fully decays after 7000 days.
In particular, we use the following $g_4$ functions:
%
\begin{eqnarray}
g_4^{\mathrm{Type\ I}}(t) &=& -83.7989\times \bigl(r\times5\times
10^{-4}+3.5\times10^{-3} \bigr)\times \frac{1}{1+e^{0.0018t}},
\\
\qquad g_4^{\mathrm{Types\ II,III,IV}}(t) &=& -49.014\times \bigl(r\times5\times
10^{-4}+7\times10^{-3} \bigr)\times \frac{1}{1+e^{0.00068t}},
\end{eqnarray}
where $r$ is randomly sampled from a standard normal distribution.
For all inspection types, we used the following $g_3$ saturation function:
\[
g_3(t) = 0.4\times \biggl(1-\log \bigl(1+e^{-3.75t} \bigr)
\times\frac{1}{\log2} \biggr),
\]
which ensures that subsequent inspections do not lower the
vulnerability to more than 60\% of the baseline vulnerability. Sampled
$g_4$ functions for Type I and Types~II, II, IV, along with $g_3$ are
shown in Figure~\ref{g3g4simulation}.


%
%
%
%

One targeted inspection per manhole was distributed randomly across $Y$
years for the bright line $Y$-year inspection policies, and $3\times
365 =1095$ ad hoc inspections for each year were uniformly distributed,
which corresponds to 3 ad hoc inspections per day for the whole power
grid on average. During the simulation, when we arrived at a time step
with an inspection, the inspection outcome was Type I with 25\%
probability, or one of Types II, III or IV, with 25\% probability. In
the rest of the cases (50\% probability), the inspection was clean, and
the manhole's vulnerability was not affected by the inspection. If the
inspection resulted in a repair, we sampled $r$ randomly and randomly
chose the inspection outcome (Type I or Types II, III, IV). This
percentage breakdown was observed approximately for a recent year of
inspections in NYC.

To initialize manhole vulnerabilities for a bright line policy of $Y$
years, we simulated the previous $Y$-year inspection cycle and started
the simulation with the vulnerabilities obtained at the end of this
full cycle.

\begin{figure}
\centering
\begin{tabular}{@{}cc@{}}

\includegraphics{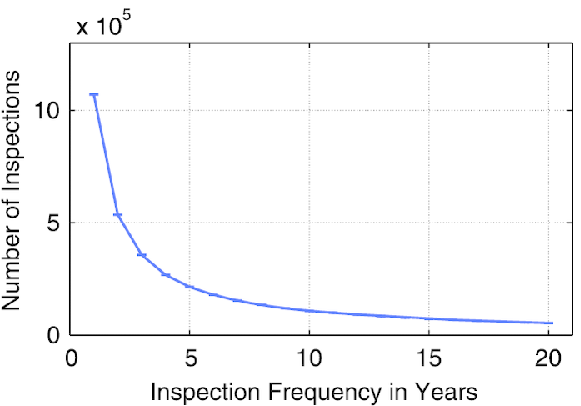}
 & \includegraphics{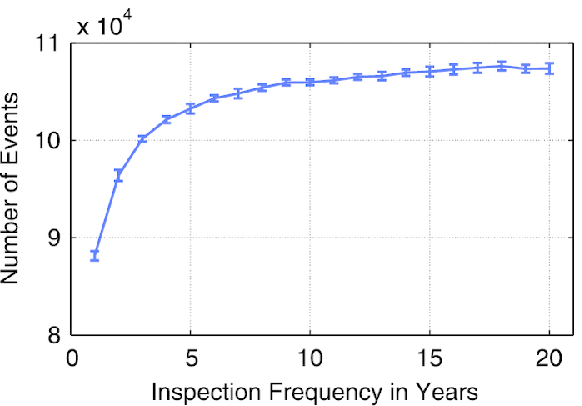}\\
\end{tabular}
\caption{Number of events and inspections based on bright line policy.
Number of years $Y$ for the bright line policy is on the horizontal
axis in both figures. The left figure shows the number of inspections,
the right figure shows the number of events.}
\label{fig:brightline_events_and_inspections}
\end{figure}

\textit{Simulation results}: We simulated events and
inspections for 53.5K manholes for bright line policies ranging from
$Y=1$ year to $Y=20$ years.
Naturally, a longer inspection cycle corresponds to fewer daily
inspections, which translates into an increase in overall
vulnerabilities and an increase in the number of events. This is
quantified in Figure~\ref{fig:brightline_events_and_inspections}, which
shows the projected number of inspections and events for each $Y$ year
bright line policy. If we change from a 6 year bright line inspection
policy to a 4 year policy, we estimate a reduction of approximately 100
events per year. The relative costs of inspections and events can thus
be considered in order to justify a particular choice of $Y$ for the
bright line policy.

Let us say, for instance, that each inspection costs $C_I$ and each
event costs $C_E$. The simulation results allow us to denote the
forecasted expected number of events over a period of time $T$ as a
function of the inspection frequency $Y$, which we denote by
$N_E(Y,T)$. The value of $N_E(Y,T)$ comes directly from the simulation,
as plotted in Figure~\ref{fig:brightline_events_and_inspections}. Let
us make a decision for $Y$ for $P$ total manholes, over a period $T$.
To do this, we would choose a $Y$ that minimizes the total cost
\[
C_E\times N_E(Y,T) + C_I\times P\times
T/Y.
\]
This line of reasoning provides a quantitative mechanism for decision
making and can be used to justify a particular choice of inspection policy.

%

\section{Conclusion}
Keeping our electrical infrastructure safe and reliable is of critical
concern, as power outages affect almost all aspects of our society,
including hospitals, financial centers, data centers, transportation
and supermarkets. If we are able to combine historical data with the
best available statistical tools, it will be possible to impact our
ability to maintain an ever aging and growing power grid. In this work,
we presented a methodology for modeling power-grid failures that is
based on natural assumptions: (i) that power failures have a
self-exciting property, which was hypothesized by Con Edison engineers,
(ii) that the power company's actions are able to regulate
vulnerability levels, (iii) that the effects on the vulnerability level
of past events or repairs can saturate, and (iv) that vulnerability
estimates should be similar between similar entities. We have been able
to show directly (using the CF estimator for the RPP) that the
self-exciting and saturation assumptions hold. We demonstrated through
experiments on past power-grid data from NYC, and through simulations,
that the RPP model is able to capture the relevant dynamics well enough
to predict power failures better than the current approaches in use.

The modeling assumptions that underlie RPPs can be directly ported to
other problems. RPPs are a natural fit for problems in healthcare,
where medical conditions cause self-excitation and treatments provide
regulation. Through the Bayesian framework we introduced, RPPs extend
to a broad range of problems where predictive power can be pooled among
multiple related entities, whether manholes or medical patients.

The results presented in this work show for the first time that manhole
events can be predicted in the short term, which was previously thought
not to be possible. Knowing how one might do this permits us to take
preventive action to keep vulnerability levels low, and can help make
broader policy decisions for power-grid maintenance through simulation
of many uncertain futures, simulated over any desired policy.


\begin{supplement}
\stitle{Supplementary material for ``Reactive point processes: A new
approach to predicting power failures in underground electrical systems''\\}
\slink[doi,text={10.1214/14-AOAS789SUPP}]{10.1214/14-AOAS789SUPP} 
\sdatatype{.pdf}
\sfilename{aoas789\_supp.pdf}
\sdescription{The supplementary material includes an expanded related
work section, conditional frequency estimator (CF estimator) for the
RPP, experiments with a maximum likelihood approach, a description of
the inspection policy used in Section~\ref{SecInspectionPolicies}, an
analysis of Manhattan data using random effects model and simulation
studies for validating the fitting techniques for the models in the
paper. It also includes a description and link for a publicly available
simulated data set that we generated, based on statistical properties
of the Manhattan data set.}
\end{supplement}

%

%

\printaddresses
\end{document}